\documentclass[11pt,twoside]{article}

\usepackage{asp2006}
\usepackage{epsf}
\usepackage{lscape}
\usepackage{graphicx}

\begin{document}
\title{First Results from the Transit Ephemeris Refinement and
Monitoring Survey (TERMS)}   

\author{Stephen R. Kane\altaffilmark{1}, Suvrath
  Mahadevan\altaffilmark{2,3}, Kaspar von Braun\altaffilmark{1}, Gregory
  Laughlin\altaffilmark{4}, Andrew Howard\altaffilmark{5}, David
  R. Ciardi\altaffilmark{1}}
\altaffiltext{1}{NASA Exoplanet Science Institute, Caltech, MS 100-22,
  Pasadena, CA 91125}
\altaffiltext{2}{Dept of Astronomy \& Astrophysics, Penn State
  University, University Park, PA 16802}
\altaffiltext{3}{Center for Exoplanets and Habitable Worlds,
  Penn State University, University Park, PA 16802}
\altaffiltext{4}{UCO/Lick Observatory, University of California, Santa
  Cruz, CA 95064}
\altaffiltext{5}{Department of Astronomy, University of California,
  Berkeley, CA, 94720}

\begin{abstract} 
Transiting planet discoveries have yielded a plethora of information
towards understanding the structure and atmospheres of extra-solar
planets. These discoveries have been restricted to the short-period or
low-periastron distance regimes due to the bias inherent in the
geometric transit probability. Through the refinement of planetary
orbital parameters, and hence reducing the size of transit windows,
long-period planets become feasible targets for photometric
follow-up. Here we describe the TERMS project which is monitoring
these host stars at predicted transit times.
\end{abstract}



\section{Introduction}

Monitoring known radial velocity planets at predicted transit times
presents an avenue through which to explore the mass-radius
relationship of exoplanets into regions of period/periastron space
beyond that which is currently encompassed \citep{kan07}. This is
particularly true for those planets in relatively eccentric orbits
\citep{kan08,kan09a}, as demonstrated by the the variation in transit
probability with the argument of periastron (see Figure 1, left
panel). Here we describe techniques for refining ephemerides and
performing follow-up observations \citep{kan09b}. These methods are
being employed by the Transit Ephemeris Refinement and Monitoring
Survey (TERMS).

\section{Transit Ephemerides}

The transit window as described here is defined as a specific time
period during which a complete transit (including ingress and egress)
could occur for a specified planet.
The size of a transit window depends upon the uncertainties in the
fit parameters. They also deteriorate over time, motivating follow-up
of the transit window as soon as possible after discovery.  Figure 1
(right panel) shows the size of the first transit window (after the
fit time of periastron passage, $t_p$) for a sample of 245
exoplanets. The transit windows of the short-period planets tend to be
significantly smaller than those of long-period planets since, at the
time of discovery, many more orbits have been monitored to provide a
robust estimate of the orbital period. Targets chosen for TERMS tend
to be those which have small transit windows, medium-long periods, and
a relatively high probability of transiting the host star.

\begin{figure*}
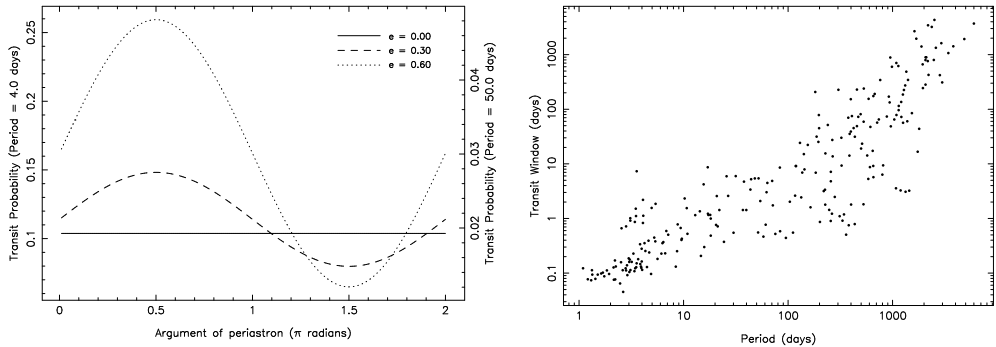

  \begin{center}
    \begin{tabular}{cc}
      \includegraphics[angle=270,width=6.5cm]{kane_proc_fig1a.eps} &
      \includegraphics[angle=270,width=6.2cm]{kane_proc_fig1b.eps} \\
    \end{tabular}
  \end{center}
  \caption{Left: Dependence of geometric transit probability on the
    argument of periastron for eccentricities of 0.0, 0.3, and 0.6,
    plotted for periods of 4.0 days and 50.0 days. Right: Ephemeris
    calculations for a sample of 245 exoplanets, showing the
    dependence of transit window size on period.}
\end{figure*}

\section{TERMS Strategy}

A considerable number of high transit probability targets are
difficult to adequately monitor during their transit windows because
the uncertainties in the predicted transit mid-points are too high
(months or even years). The acquisition of just a handful of new
radial velocity measurements at carefully optimised times can reduce
the size of a transit window by an order of magnitude. This is
described in more detail by \citet{kan09b}.

Due to the wide range of stars monitored, both in sky location and
brightness, TERMS collaborates with a variety of existing groups to
take advantage of transit window opportunities. Note that the
observations from this survey will lead to improved exoplanet orbital
parameters and ephemerides even without an eventual transit detection
for a particular planet. The results from this survey will provide a
complementary dataset to the fainter magnitude range of the {\it
  Kepler} mission, which is expected to discover many transiting
planets including those of intermediate to long-period.



\end{document}